\documentclass[aps,prl,twocolumn,superscriptaddress]{revtex4-2}

\usepackage[english]{babel}
\usepackage{amssymb,amsmath}
\usepackage{graphicx}

\begin{document}

\title{Coherent and incoherent tunneling into YSR states revealed by atomic scale shot-noise spectroscopy}

\author{U. Thupakula}
\affiliation{Laboratoire de Physique des Solides (CNRS UMR 8502), B\^{a}timent 510, Universit\'{e} Paris-Sud/Universit\'{e} Paris-Saclay, 91405 Orsay, France}

\author{V. Perrin}
\affiliation{Laboratoire de Physique des Solides (CNRS UMR 8502), B\^{a}timent 510, Universit\'{e} Paris-Sud/Universit\'{e} Paris-Saclay, 91405 Orsay, France}

\author{A. Palacio-Morales}
\affiliation{Laboratoire de Physique des Solides (CNRS UMR 8502), B\^{a}timent 510, Universit\'{e} Paris-Sud/Universit\'{e} Paris-Saclay, 91405 Orsay, France}

\author{L. Cario}
\affiliation{Universit\'{e} de Nantes, CNRS, Institut des Mat\'{e}riaux Jean Rouxel, IMN, F-44000 Nantes, France}

\author{M. Aprili}
\affiliation{Laboratoire de Physique des Solides (CNRS UMR 8502), B\^{a}timent 510, Universit\'{e} Paris-Sud/Universit\'{e} Paris-Saclay, 91405 Orsay, France}

\author{P. Simon}
\affiliation{Laboratoire de Physique des Solides (CNRS UMR 8502), B\^{a}timent 510, Universit\'{e} Paris-Sud/Universit\'{e} Paris-Saclay, 91405 Orsay, France}

\author{F. Massee}
\email[]{freek.massee@universite-paris-saclay.fr}
\affiliation{Laboratoire de Physique des Solides (CNRS UMR 8502), B\^{a}timent 510, Universit\'{e} Paris-Sud/Universit\'{e} Paris-Saclay, 91405 Orsay, France}

\date{\today}

\begin{abstract}
The pair breaking potential of individual magnetic impurities in s-wave superconductors generates localized states inside the superconducting gap commonly referred to as Yu- Shiba-Rusinov (YSR) states whose isolated nature makes them ideal building blocks for artificial structures that may host Majorana fermions. One of the challenges in this endeavor is to understand their intrinsic lifetime, $\hbar/\Lambda$, which is expected to be limited by the inelastic coupling with the continuum thus leading to decoherence. Here we use shot-noise scanning tunneling microscopy to reveal that electron tunnelling into superconducting 2H-NbSe$_2$ mediated by YSR states is ordered as function of time, as evidenced by a reduction of the noise. Moreover, our data show the concomitant transfer of charges e and 2e, indicating that incoherent single particle and coherent Andreev processes operate simultaneously. From the quantitative agreement between experiment and theory we obtain $\Lambda$ = 1~$\mu$eV $\ll$ $k_BT$ demonstrating that shot-noise can probe energy- and time scales inaccessible by conventional spectroscopy whose resolution is thermally limited.
\end{abstract}

\maketitle

\section{Introduction}
For a single impurity spin in a superconductor, a bound state appears inside the gap \cite{yu, shiba, rusinov} with its particle- and hole components located in energy symmetrically around zero \cite{balatsky_revmodphys_2006}. The amplitudes of the two components are usually different \cite{yazdani_science_1997, ji_prl_2008, heinrich_prog_2018} and reflect the coherence factors of the electron and hole excitations. Whereas the spatial extent of the YSR states is typically on the order of a few atoms \cite{yazdani_science_1997}, they have been shown to range up to tens of nanometers for two dimensional superconductors \cite{menard_nphys_2015}. Since the YSR states are inside the superconducting gap, one of the key questions, particularly with an eye on building more complicated structures such as chains and islands \cite{nadj_science_2014, ruby_prl_2015a, pawlak_npjqi_2016, kim_sciadv_2018} is how electrons move between the YSR states and the condensate. Not only is a full understanding of the nature of the impurity and its intrinsic lifetime important for theoretical modelling, the dynamics of charge transfer through it can in principle also be used to directly distinguish conventional YSR states from Majorana bound states \cite{vivien_prb}. 

\begin{figure}
	\centering
	\includegraphics[width=0.8\columnwidth]{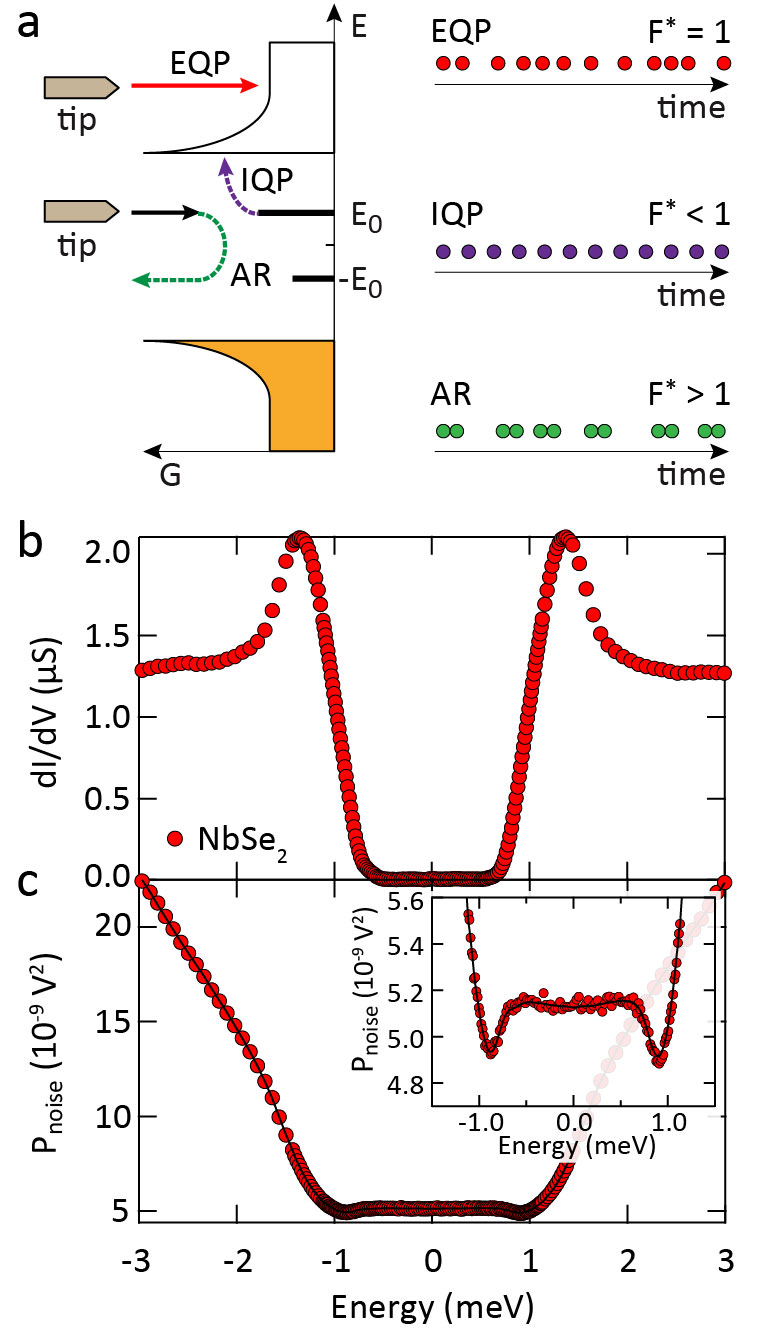}
	
	\caption{\label{fig:1} \textbf{Noise schematic and fully gapped 2H-NbSe$_2$}. \textbf{a} Schematic depiction of the various tunnelling processes into 2H-NbSe$_2$: direct elastic quasi-particle tunnelling into the bulk (EQP, red) inelastic quasi-particle tunnelling through the YSR state (IQP, purple) and Andreev reflection assisted by the YSR (AR, green). Whereas direct tunnelling of electrons generates Poissonian noise ($F^*=1$), inelastic tunneling through the YSR state orders the electron flow, reducing the noise ($F^*<1$). The energy relaxation time sets the intrinsic lifetime of the YSR. Andreev reflection on the other hand involves a charge of q = 2e, leading to $F^*>1$. The coloured spheres for the three scenarios on the right illustrate the passage of electrons as function of time: random, ordered, or in pairs. \textbf{b} Differential conductance on a fully gapped 2H-NbSe$_2$ surface, $V$ = 3~mV, $I$ = 3~nA. \textbf{c} Current noise power recorded simultaneously with \textbf{b}, the black line shows a fit for $F^* = 1$, that takes into account the measured $R_J(V)$ and uses a single set of fitting parameters for all voltages. The inset highlights the effect of the strongly varying dynamical resistance on the measured voltage noise which is accurately captured by the fit. See Supplementary Information section 1 for more details.}
\end{figure}

The high spatial- and energy resolution make the scanning tunnelling microscope (STM) an ideal experimental tool to investigate the tunnelling process into individual impurities. Previous studies using the time averaged current and theoretical modelling showed tantalizing signatures of a transition from single-electron dominated tunnelling to multi-particle Andreev processes \cite{ruby_prl_2015, huang_nphys_2020}. Direct evidence for such a transition, and more specifically for Andreev processes to occur with a standard metallic probe, however, is lacking. Moreover, direct tunnelling into the magnetic impurity can complicate the interpretation of the data, since the impurity state may be affected by the presence of the tip \cite{heinrich_nanoletters_2015, ormaza_ncomm_2017, malavolti_nanoletters_2018, liljeroth_nanoletters_2019}, or multiple tunnelling paths and relaxation processes may occur \cite{figgins_prl_2010, bryant_nanoletters_2015, farinacci_prl_2020} which are difficult to incorporate in theory. Although experimentally challenging, atomic scale shot-noise measurements can in principle resolve all these issues. This is because shot-noise ($S_{shot}$), which is current noise originating from the discreetness of the charge carriers, is sensitive to both the charge of the carriers, $q$, as well as the nature of the tunnelling process \cite{blanter_buttiker}. To evaluate the shot-noise, we consider the effective Fano factor, $F^* = S_{shot}/2e|I| = qF/e$, where $I$ is the time averaged current, $e$ the electron charge and $F$ the Fano factor. For uncorrelated elastic quasi-particle tunnelling between a metallic tip and the bulk superconductor, the electrons follow Poissonian statistics \cite{ronen_pnas} giving $F^*=F=1$ (EQP in Fig. \ref{fig:1}a). Single electron tunnelling via the YSR states, on the other hand, is a resonant process and only possible through inelastic quasi-particle relaxation (IQP in Fig. \ref{fig:1}a). Therefore the electron flow will become ordered, thereby reducing the current noise from Poissonian to sub-Poissonian, i.e. $F^*<1$ (IQP in Fig. \ref{fig:1}a), as observed in quantum dots \cite{choi_apl}. Alternatively, Andreev reflection through the YSR state transfers a charge of $q = 2e$, giving $F^*>1$ (AR in Fig. \ref{fig:1}a), enabling them to be easily distinguished from either forms of single electron tunnelling \cite{jehl_nature, ronen_pnas, bastiaans_prb}.

\section{Results}
Noise at the atomic scale is measured using our home-built scanning tunnelling microscope with cryogenic circuitry operating in the MHz regime \cite{massee_rsi_2018}. The circuitry, which is explained in more detail in the Supplementary Information, converts the current noise into voltage noise at the input of a cryogenic amplifier and can operate simultaneously with conventional low frequency (DC) measurements. Although in most circumstances the junction resistance, $R_J$, is sufficiently large to be safely ignored, for the relatively low junction resistances used in this work it will start acting as a voltage divider resulting in a reduction of the measured voltage noise. Importantly, for a highly non-linear $I(V)$ characteristic, such as that of clean 2H-NbSe$_2$ (Fig. \ref{fig:1}b), the dynamical resistance, $R_J(V)=dV/dI$, needs to be considered instead of the setup resistance. Fig. \ref{fig:1}c shows this effect vividly: instead of a simple linear current dependence of the measured noise power, it decreases at the onset of the coherence peak ($\sim\pm 0.9~mV$) below that obtained at zero voltage. As evidenced by the fit in Fig. \ref{fig:1}c, this non-linear behavior is purely an effect of the non-linearity of the junction, $R_J(V)$, which reduces the contribution of the thermal noise more than that of the shot-noise increases; $F^*=1$ for all voltages as expected for quasi-particle tunnelling following Poissonian statistics \cite{ ronen_pnas} (EQP in Fig. \ref{fig:1}a).

\begin{figure}
	\centering
	\includegraphics[width=0.85\columnwidth]{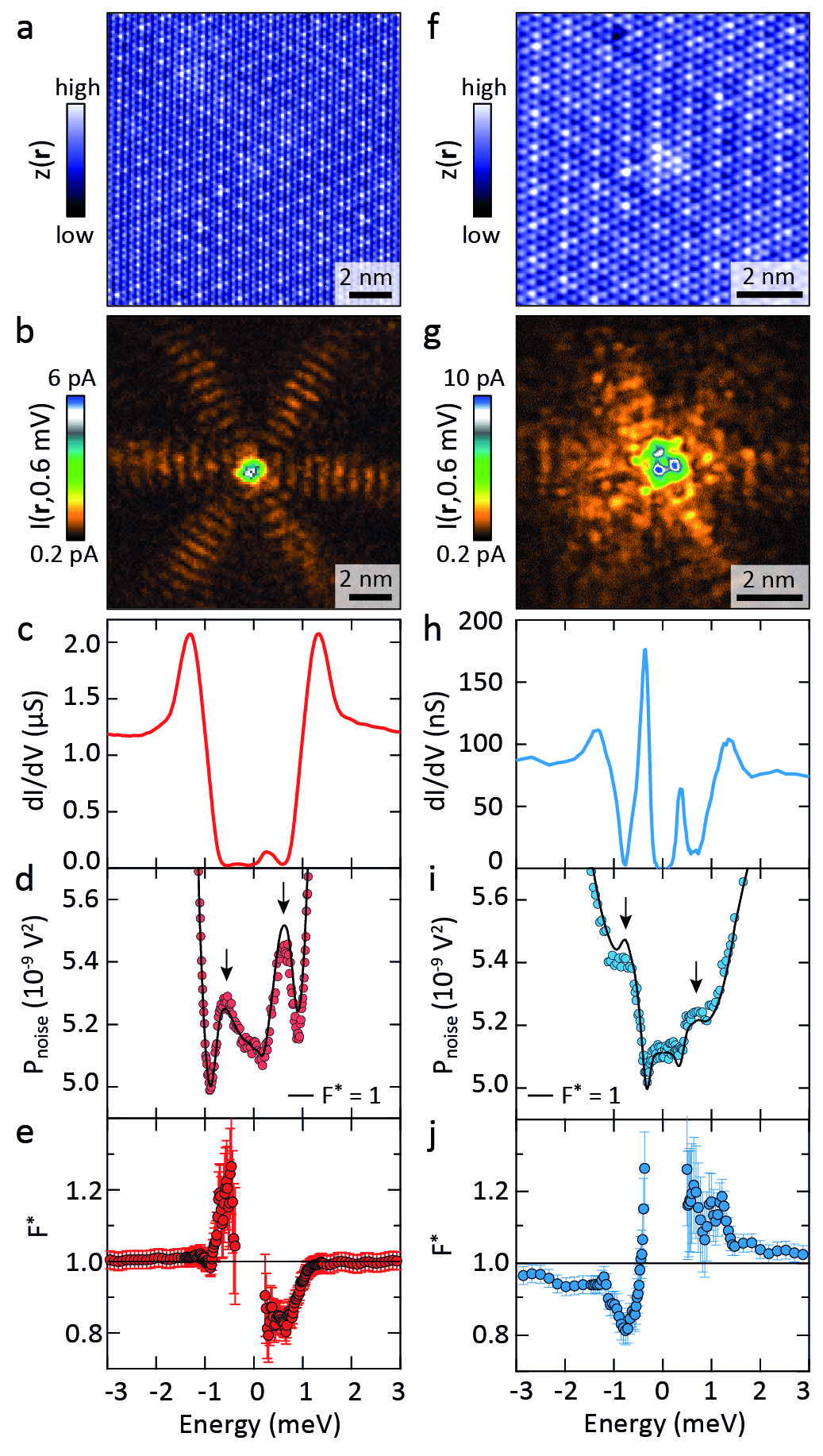}
	
	\caption{\label{fig:2} \textbf{YSR cores in 2H-NbSe$_2$}. \textbf{a} Atomically resolved constant current image of the Se surface of 2H-NbSe$_2$, $V$ = 4.2~mV, $I$ = 100~pA. \textbf{b} In-gap current recorded simultaneously with \textbf{a} showing a spatially extended YSR states generated by a sub-surface impurity. \textbf{c} Tunnelling spectrum taken at the core of the YSR and \textbf{d} simultaneously recorded current noise, $V$ = 3~mV, $I$ = 3~nA. The black line in \textbf{d} indicates $F^* = 1$: the noise from the YSR deviates at both positive and negative sample bias (arrows), as can be clearly seen in the effective Fano factor, $F^*$, in \textbf{e}. Outside the gap, $F^*$ converges to 1. \textbf{f}-\textbf{j} The same as \textbf{a}-\textbf{e} for a more compact, but relatively strong, YSR located elsewhere on the same sample. $V$ = 4~mV, $I$ = 100~pA for \textbf{f}, \textbf{g}; $V$ = 6~mV, $I$ = 400~pA for \textbf{h}-\textbf{j}.}
\end{figure}

Having a detailed understanding of the noise data recorded on clean 2H-NbSe$_2$, we next move to the electronic transport into YSR states in 2H-NbSe$_2$. The precise spatial extent of the in-gap resonances, as well as the particle-hole asymmetry of the core, can vary from impurity to impurity as it depends on the impurity type and its location in the crystal structure \cite{menard_nphys_2015, senkpiel_prb_2019, liebhaber_nanoletters_2020}. Two YSR states with opposite particle-hole asymmetry and different spatial extent are shown in Fig. \ref{fig:2}. Interestingly, also the ratio of the YSR- and coherence peaks is very different for the two cores. As we are interested in the tunnelling process through the YSR states, to facilitate comparison we choose setup parameters that give roughly the same YSR-current, i.e. the YSR peaks in differential conductance in Figs. \ref{fig:2}c and h have approximately the same height. Consequently, since the noise is proportional to the current, the simultaneously recorded current noise has a similar magnitude for the two cores, despite the rather different conditions outside the gap, see Figs. \ref{fig:2}d, i. To evaluate the in-gap noise, we fit the normal state ($|V|>$ 1.2~mV) noise data using the measured $R_J(V)$ and $F^*=1$, then extend the fit inside the gap giving the black line in Figs. \ref{fig:2}d, I. Unlike for the noise taken on clean 2H-NbSe$_2$ (Fig. \ref{fig:1}c), the noise on the YSR cores evidently does not follow $F^*$ = 1: for the dominant resonance, i.e. positive bias for Figs. \ref{fig:2}c, d and negative bias for Figs. \ref{fig:2}h, i, the noise is reduced ($F^*<1$), whereas for the weak resonance, i.e. negative bias for Figs. \ref{fig:2}c, d and positive bias for Figs. \ref{fig:2}h, i, it is enhanced ($F^*>1$). Although the deviations in absolute numbers are relatively small, they are larger than the experimental error bars due to our high signal-to-noise ratio and accurate fitting procedure (see Supplementary Information section 1), and correspond to changes in $F^*$ of the order of 10\% (Figs. \ref{fig:2}e, j). Importantly, the observation of $F^*>1$ strongly suggests Andreev processes to be present, as it cannot occur for single electron tunnelling into a one-level system. Furthermore, the particle-hole asymmetry of the noise also suggests a contribution from single electron tunneling, meaning both processes operate simultaneously.

\begin{figure}
	\centering
	\includegraphics[width=0.7\textwidth]{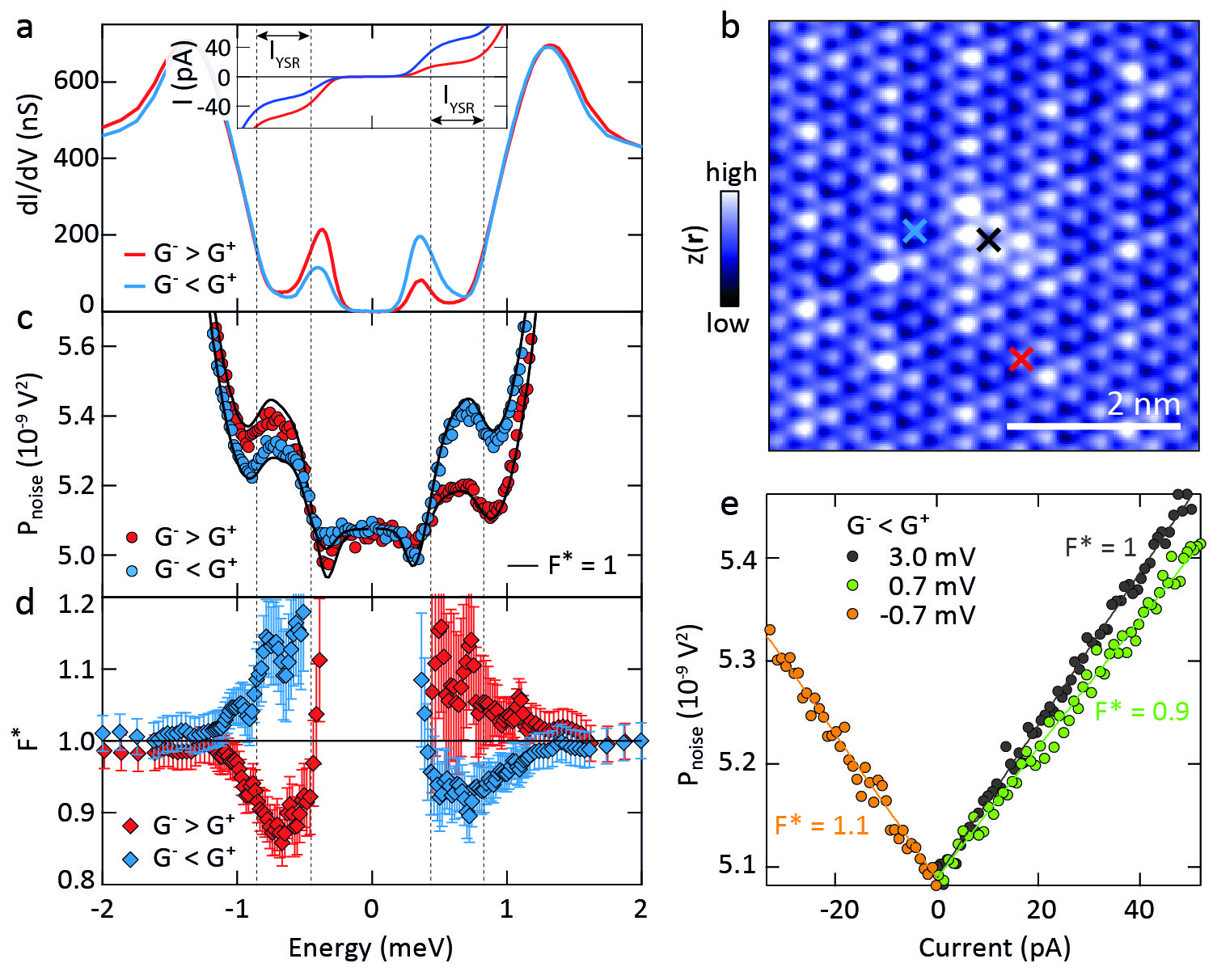}
	
	\caption{\label{fig:3} \textbf{Noise on the YSR tails}. \textbf{a} Differential conductance $G=dI/dV$ on two tail locations with opposite particle-hole asymmetry, $V$ = 4.2~mV, $I$ = 1.5~nA. We define $G^{\pm}=G(\pm E_0)$ with $E_0$ the YSR resonance energy. The inset shows the corresponding currents on the same energy scale, dashed lines indicate where the current is dominated by the YSR and is sufficiently large for noise measurements. The YSR state is the same as that in Fig. \ref{fig:2}f-j. The location of the spectra (red and blue crosses), as well as that of the YSR core (black cross) is marked in the constant current image in \textbf{b} ($V$ = 4~mV, $I$ = 100~pA). \textbf{c} Current noise recorded simultaneously with \textbf{a}, the black lines indicate $F^* = 1$. \textbf{d} Effective Fano factor, $F^*$, extracted from \textbf{c} showing clearly that the noise deviates from $F^*=1$ where the current is dominated by the YSR. The points corresponding to I$<$10~pA have been omitted for clarity; since the fluctuations in the noise are the same for all voltages, the associated error in $F^*$ becomes relatively large for such small currents. \textbf{e} Current dependence of the noise at fixed voltage taken at the blue cross in \textbf{d}.}
\end{figure}

To make a more quantitative analysis of the noise, and in order to avoid possible mechanical \cite{heinrich_nanoletters_2015, ormaza_ncomm_2017, malavolti_nanoletters_2018, liljeroth_nanoletters_2019}, multi-paths related \cite{figgins_prl_2010, bryant_nanoletters_2015, farinacci_prl_2020}, or spin-dependent \cite{pradhan_prl_2018} complications from direct tunnelling into the impurity, we shift our attention to the YSR tails. Since the particle-hole asymmetry of the tails oscillates as function of distance from the core \cite{menard_nphys_2015}, we can probe locations where the particle-hole asymmetry is nearly perfectly mirrored while all other experimental parameters remain identical. Fig. \ref{fig:3} shows two such locations, both situated a distance of several atoms from the core (crosses in Fig. \ref{fig:3}b), ensuring that direct tunnelling into the impurity is negligible. Fig. \ref{fig:3}c shows the noise recorded simultaneously with the tunnelling spectra of Fig. \ref{fig:3}a. As was the case for the spectra taken on the core, the noise is reduced ($F^*<1$) for the dominant resonance, and enhanced ($F^*>1$) for the weaker resonance. To analyse the deviation from $F^*=1$ in more detail, we extract $F^*$ for each voltage in Fig. \ref{fig:3}d which shows that $F^*$ is roughly equally enhanced as it is suppressed, and converges to Poissonian as soon as the current becomes dominated by the quasi-particles at the coherence peaks. To confirm that the noise deviates from $F^*=1$, we additionally measured the current dependence of the noise at fixed voltage in- and outside the gap, see Fig. \ref{fig:3}e.

\begin{figure}
	\centering
	\includegraphics[width=0.7\textwidth]{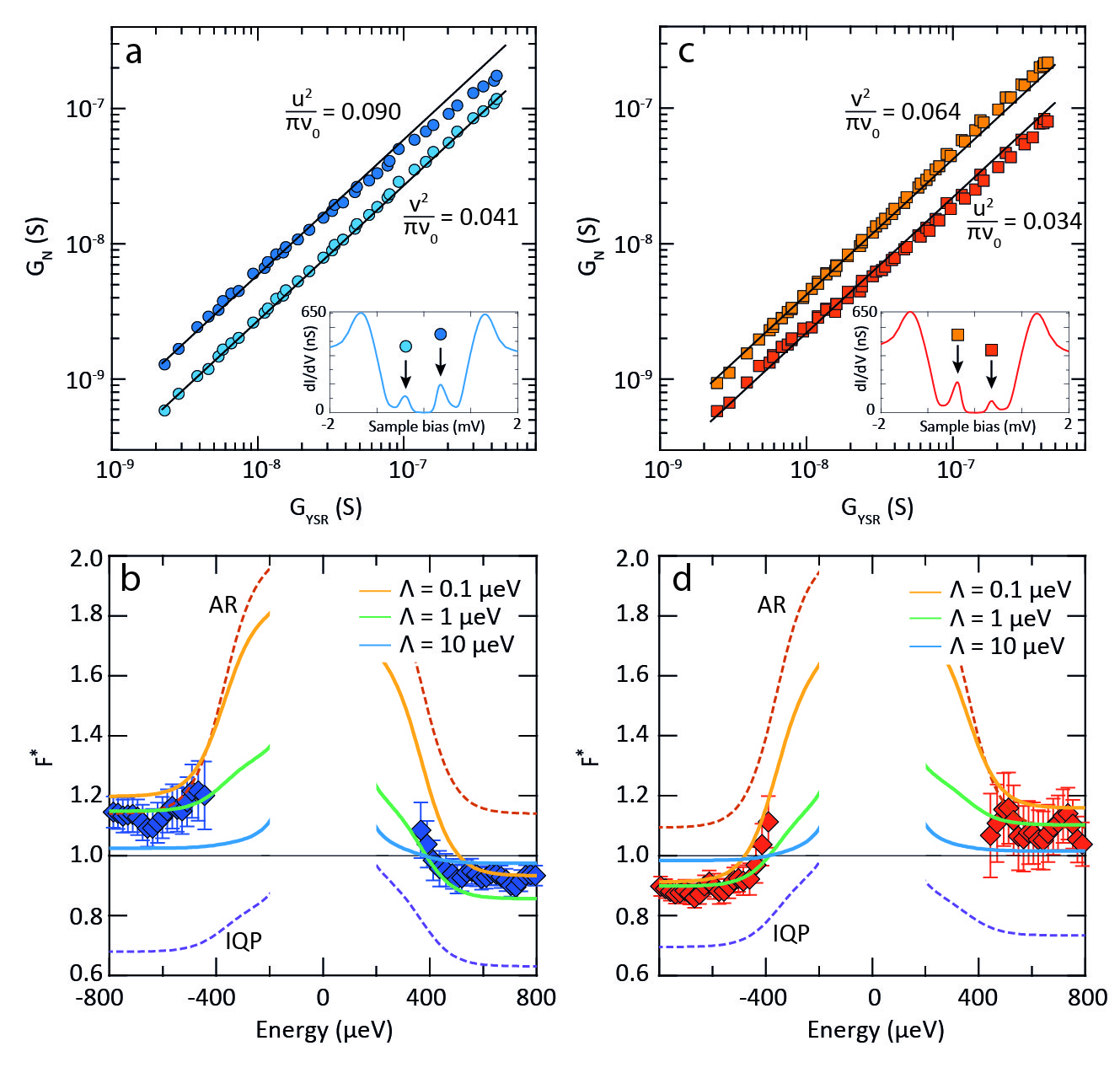}
	
	\caption{\label{fig:4} \textbf{Experiment vs theory}. \textbf{a} YSR peak conductance ($G_{YSR}$) as function of normal state conductance ($G_N$) for the location of the blue spectrum of Fig. \ref{fig:3}a, reproduced here in the inset, see Supplementary Information section 2 for more details. A linear fit of the data taken at small conductance is used to extract $u$ and $v$. \textbf{b} Sub-gap $F^*$ (symbols) for the blue spectrum of Fig. \ref{fig:3}d and theoretical curves for three different values of $\Lambda$. Pure Andreev reflection (AR, $\Lambda = 0 ~\mu eV$) and pure inelastic quasi particle tunnelling (IQP, $\Lambda = 1 ~\mu eV$) are shown for comparison. \textbf{c}, \textbf{d} Same as \textbf{a}, \textbf{b} for the red data of Fig. \ref{fig:3}.}
\end{figure}

To gain more insight into our experimental data, we calculate the tunnelling current and current noise using a standard description of a classical YSR impurity (see Supplementary Information section 2 for more details), considering both single electron tunnelling and Andreev processes \cite{ruby_prl_2015}. The main input parameters of the theory are the coherence factors of the electron and hole excitations of the YSR, $u$ and $v$, the superconducting gap, $\Delta$, the YSR energy, $E_0$, the normal state conductance, $G_N$, and the intrinsic lifetime, $\tau= \hbar/\Lambda$. Of these, $\Delta$, $E_0$ and $G_N$ can be directly obtained from the differential conductance spectra (e.g Fig. \ref{fig:3}a). Furthermore, $u$ and $v$, can be extracted from the normal state conductance ($G_N$) dependence of the YSR conductance ($G_YSR$), see Figs. \ref{fig:4}a, c. Here, we fit the single electron tunnelling dominated linear part at low conductance \cite{ruby_prl_2015, huang_nphys_2020} to determine $u$ and $v$, see also Supplementary Information section 2. This enables us to calculate the noise with as only adjustable parameter the intrinsic lifetime, $\hbar/\Lambda$, which is masked in experiment by thermal broadening.

\onecolumngrid

\twocolumngrid

Figs. \ref{fig:4}b, d compare the experimental noise data for energies where the tunnelling current is carried exclusively by the YSR states (i.e. $|V| <$ 0.8~mV) with theoretical curves for several values of $\Lambda$, showing quantitative agreement for $\Lambda\sim 1 \mu$eV, i.e. $\tau \sim$ 0.7~ns, for both measured particle-hole asymmetries. We stress that for increasing values of $\Lambda$, meaning shorter relaxation times, single electron tunnelling becomes less resonant (i.e. less ordered in time) and increasingly dominates the tunnelling process. Indeed, for $\Lambda =$ 10~$\mu$eV, the calculated noise is already nearly Poissonian ($F^*=1$) for all voltages (see Figs. \ref{fig:4}b, d), putting a strong lower limit on the relaxation time extracted from the noise data. For comparison, calculations using only Andreev reflection (AR) or only inelastic single electron processes (IQP) fail to reproduce the data for any $\Lambda$, as the former always has $F^*>1$ for both polarities and the latter $F^*<1$.

\section{Discussion}
The persistent enhancement of the noise for the smaller of the two YSR resonances in both experiment and theory directly proves the presence of Andreev processes, whereas the reduction in noise of the strong resonance results from a finite lifetime of the YSR. The fact that we obtain qualitative agreement between theory and experiment for $u$ and $v$ extracted from the linear part of Figs. \ref{fig:4}a, c, while all experimental noise data was recorded in the non-linear part of Figs. \ref{fig:4}a, c highlights the robustness of our results. We stress that although qualitatively still in agreement, the theoretical curves for the data taken on the cores (Figs. \ref{fig:4}d, h) slightly deviate quantitatively (see Supplementary Information Fig. S7), suggesting that tip and/or current induced effects may indeed play a role on the YSR core which is currently not included in the theory. Despite the less accurate fit, the noise recorded for both cores agrees best to the theory curves that use a lifetime similar to that obtained on the YSR tails. This implies that for roughly equal YSR currents, which at sub-gap voltages constitute the only contribution to the current, the YSR lifetime in 2H-NbSe$_2$ is independent of the particular details of the magnetic impurity and the measurement location. Furthermore, the sub-nanosecond value at 0.7~K we extract from the noise data is similar to that reported for Mn atoms on a Pb(111) surface \cite{ruby_prl_2015}, and not incompatible with the electron-phonon relaxation time in 2H-NbSe$_2$ \cite{anikin_prb_2020}. Importantly, the energy scale of relaxation we obtain ($\Lambda = 1~\mu$eV) is much smaller than the electron temperature, meaning that shot-noise allows to determine the YSR lifetime even if the width of the resonance in the tunnelling spectra is dominated by thermal broadening, or the current not thermally saturated as used previously \cite{ruby_prl_2015}. More generally, we show that local measurements of fluctuations in the tunnelling current provide direct evidence of the coherent transfer of charges equal to $2e$ through YSR states into the bulk, and enable one to probe energy- and time scales inaccessible to conventional spectroscopy. Although experimentally challenging, this work shows the feasibility of using atomic scale shot-noise to elucidate the transport dynamics through individual impurity resonances, which could in future studies be used to e.g. distinguish trivial from non-trivial states.

\begin{acknowledgments}
We thank A. Mesaros and M. Civelli for insightful discussions. We would like to acknowledge funding from H2020 Marie Sk\l{}odowska-Curie Actions (grant number 659247) and the ANR (ANR-16-ACHN-0018-01 and ANR-19-CE47-0006). 
\end{acknowledgments}

\end{document}